\documentclass[]{interact}

\usepackage{epstopdf}
\usepackage[caption=false]{subfig}
\pdfoutput=1
\usepackage[numbers,sort&compress]{natbib}
\bibpunct[, ]{[}{]}{,}{n}{,}{,}
\makeatletter
\def\NAT@def@citea{\def\@citea{\NAT@separator}}
\makeatother

\theoremstyle{plain}

\theoremstyle{definition}

\theoremstyle{remark}

\begin{document}


\title{Machine Learning for Quantum Matter}

\author{
\name{Juan Carrasquilla\textsuperscript{ }\thanks{ Email: carrasqu@vectorinstitute.ai}} 
\affil{\textsuperscript{}Vector Institute for Artificial Intelligence, MaRS Centre, Toronto, ON, Canada M5G 1M1 and Department of Physics and Astronomy, University of Waterloo, Ontario, N2L 3G1, Canada} 
}

\maketitle

\begin{abstract}
  Quantum matter, the research field studying phases of matter whose properties are
  intrinsically quantum mechanical, draws from areas as diverse as hard condensed
  matter physics, materials science, statistical mechanics, quantum information,
  quantum gravity, and large-scale numerical simulations. Recently, researchers interested
  quantum matter and strongly correlated quantum systems have turned their attention to 
  the algorithms underlying modern machine learning with an eye on making progress in 
  their fields. Here we provide a short review on the recent development and adaptation 
  of machine learning ideas for the purpose advancing research in quantum  matter, including 
  ideas ranging from algorithms that recognize conventional and topological states of matter 
  in synthetic an experimental data, to representations of quantum states in terms of 
  neural networks and their applications to the simulation and control of quantum systems. 
  We discuss the outlook for future developments in areas at the intersection between 
  machine learning and quantum many-body physics.
\end{abstract}

\begin{keywords}
Strongly correlated quantum systems; machine learning
\end{keywords}

\section{Introduction}
Machine learning studies algorithms and statistical models that computers use to perform
tasks without explicit instructions~\cite{Bishop2006}. Machine learning technology currently powers an
ever-growing number of aspects of our society including web search, virtual personal assistants,
traffic predictions, face recognition on social networks, content filtering on
commerce websites, email spam filtering, language translation, online fraud detection,
and more~\cite{LeCun2015,Goodfellow-et-al-2016}. The origin behind these technological
advances can be largely traced back to a series of breakthroughs in artificial intelligence,
in particular those based on deep learning, where data is processed through the sequential
combination of multiple nonlinear layers~\cite{Goodfellow-et-al-2016}. Deep learning has
accelerated the adoption of artificial intelligence with notable advances in areas ranging
from computer vision~\cite{voulodimos2018} and
natural language processing~\cite{young2018}, to scientific applications such as drug
discovery\cite{vamathevan2019} and protein folding~\cite{senior2020}.

Recently, the condensed matter physics, quantum information, statistical physics, and atomic,
molecular, and optical physics communities have turned their attention to the algorithms
underlying modern machine learning with the objective of making progress in quantum matter
research. This recent resurgence of research interest at the intersection
between strongly correlated systems and machine learning is shaped in part by the
commonalities in the structure of the problems that these seemingly unrelated fields attack.
For example, the complexity associated with the study of the collective behaviour of many-body
systems is reflected in the size of the state space, which grows exponentially with the
number of particles. Likewise, the ``curse of dimensionality''~\cite{10.5555/862270,Bishop2006}
affects, e.g., computer vision and natural language processing, where the size of the space
where images and sentences live grows exponentially with the number of pixels and words,
respectively. Beyond high dimensionality, many-body systems, as well as systems of interest
in machine learning, exhibit correlations and symmetries with strikingly similar structure.
A prominent example of this appears in natural language~\cite{Ebeling_1994},
natural images~\cite{ruderman1994}, and music~\cite{Levitin3716},
all of which exhibit power-law decaying correlations identical to a (classical or quantum)
many-body system tuned at its critical point, which also exhibits spatial correlations 
decaying with a power law~\cite{sachdev1999quantum}. In the same vein, the adoption of 
a common set of symmetries simultaneously enriches our understanding of quantum 
systems~\cite{greiner1989quantum} and simplifies the computational and sample complexity 
of certain learning tasks~\cite{NIPS2014_5424,10.555530453903045705}.

All these common structures suggest that the power and scalability of modern machine
learning architectures and algorithms are naturally well-suited to applications in
physical systems, in particular to help perform various tasks in strongly correlated
systems, quantum matter, quantum information and computation, and statistical physics. 
Here we review recent advances in the development and adaptation of machine learning techniques
for the purpose advancing research in these areas. This short review starts with a very
brief introduction to the essential ideas in machine learning used throughout the review but it
does not explain the technical and methodological details. An exceptional discussion of machine
learning is available in Ref.~\cite{mehta2019}, where an introduction to the core
concepts and tools of machine learning is presented from a physicists' viewpoint. Comprehensive
and extremely clear books dealing with modern machine learning ideas include Bishop's Pattern Recognition
and Machine Learning~\cite{Bishop2006}, The Elements of Statistical Learning~\cite{hastie_09_elements-of.statistical-learning},
and Deep Learning~\cite{Goodfellow-et-al-2016}. This review is focused on machine learning 
applications to quantum matter and strongly correlated many-body systems, which are physical systems 
where interactions play a major role in determining their properties. Examples of such 
strongly interacting systems include conventional and high-temperature superconductors, 
magnetic systems, quantum Hall systems, systems of electrons in confined one- and two-dimensional 
geometries, electronic and bosonic insulating states, among many others~\cite{avella2012}. 
A review dealing with the recent research at the interface between machine learning and the broader physical sciences
is available in Ref.~\cite{RevModPhys.91.045002}. Another related research domain that is not
covered here is quantum machine learning, which refers to  the development and use
of machine learning algorithms on quantum devices; this is reviewed in Ref.~\cite{Biamonte2017,Dunjko_2018}. 
Peter Wittek, who sadly disappeared during an expedition on Mount Trishul, and Vedran Dunjko wrote
a non-review of quantum machine learning~\cite{Dunjko2020nonreviewofquantum} where they provide
a perspective on the meaning of quantum machine learning, its key issues, progress, and recent trends.   
A very thorough review with a strong focus on the applications of machine learning in solid-state
materials science can be found in Ref.~\cite{schmidt2019}. A review at the intersection between
the broad area of quantum physics and artificial intelligence~\cite{Dunjko_2018} 
discusses a growing body of recent work at the intersection between quantum computing and machine
learning and how results and techniques from one field can be used to tackle the problems of
the other. This includes quantum computing as a means to provide speed-ups for machine learning 
problems, machine learning for advancing quantum technology, and quantum generalizations of 
statistical learning concepts.  

After a brief introduction to the main ideas in machine learning, this review is structured according
to the following intertwined research trends:
\begin{itemize}
  \item Machine learning in simulations of strongly correlated fermions 
  \item Machine learning phases of matter in simulated and experimental data
  \item Neural-network quantum states and their applications
  \item Machine learning acceleration of Monte Carlo simulations
  \item Quantum information, quantum control, and quantum computation
  \item Quantum physics inspired machine learning
\end{itemize}

\section{Machine learning}

The field of artificial intelligence deals with the theory and development of
computational systems endowed with the ability to perform tasks that typically
require human capabilities such as visual and speech recognition, language comprehension,
decision making, etc. The field of artificial intelligence has already solved a wide array of
problems that are laborious for human beings but straightforward for computers. The
solutions to this breed of problems can be described by a list of formal rules that
computers can process efficiently. Modern machine learning, instead, deals in part
with the challenge of automatizing the solution of tasks that are in principle easy
for humans but that are hard to formally describe by simple rules. Thus, machine learning
can be understood as a sub-field of artificial intelligence which studies algorithms,
software, and statistical models to automate tasks without explicit instructions.

Machine learning algorithms are typically divided into the categories of supervised,
semi-supervised, unsupervised, and reinforcement learning. As we will discuss, algorithms
belonging to all these categories have been applied to quantum systems. While for some
of these machine learning categories there are no formal differences when described in the language of
probability, such a division is often useful as a way to specify the details of the
algorithms, the training setup, and the structure of the datasets involved in the
learning task.

\subsection{Supervised Learning}
\begin{figure}
\centering
{%
\resizebox*{9cm}{!}{\includegraphics{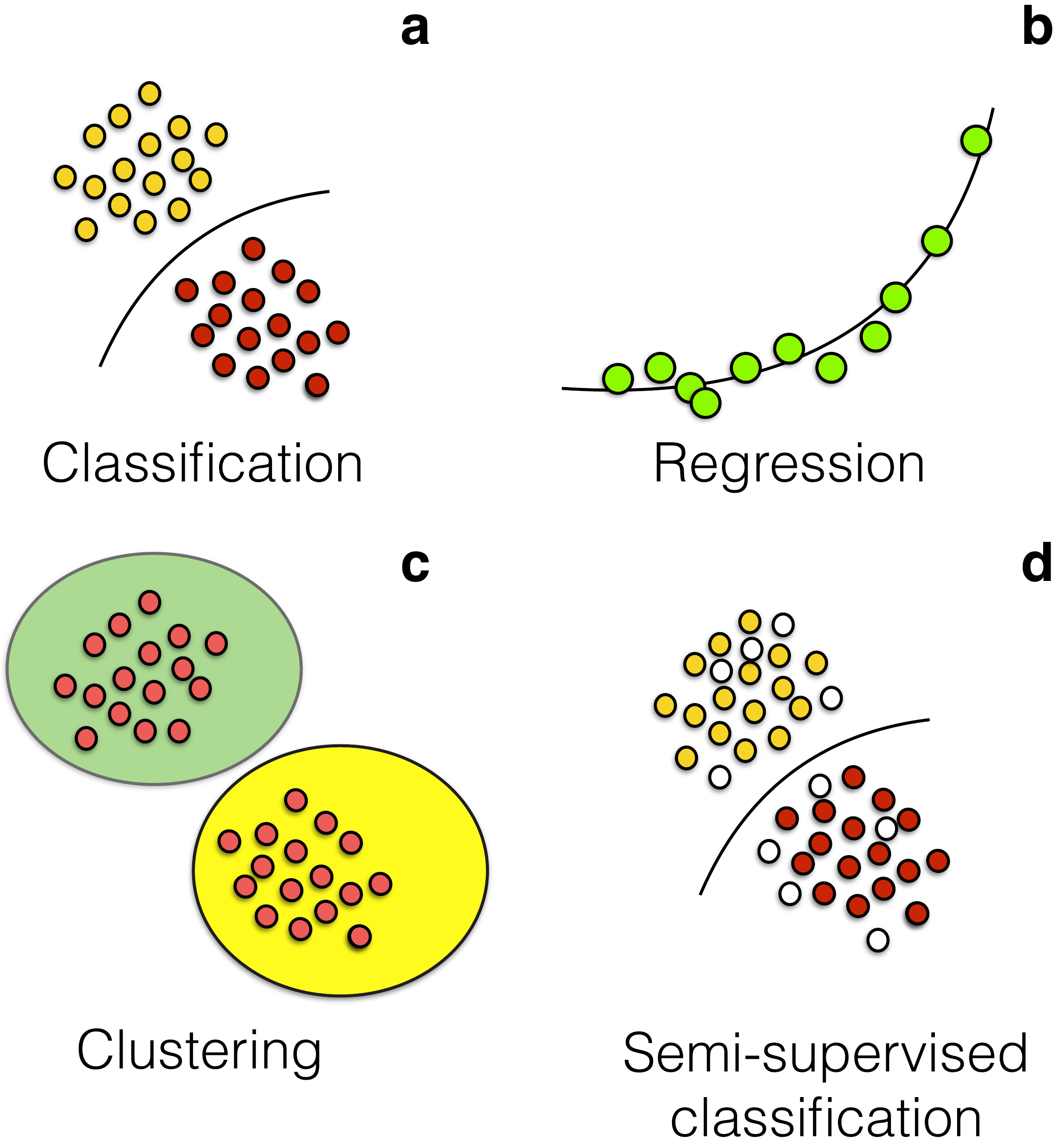}}}\hspace{5pt}
\caption{Illustrating the different categories of machine learning tasks. a. In classification,
each learning example is associated with a discrete category or target value,
which corresponds to a class. There can be, e.g., two classes
in binary classification (red and yellow). The function separating the two
classes is called decision boundary (black curve). b. In a regression each
learning example is associated with a real target value. The goal of the model
(black curve in the figure) is to estimate the correct output, given an input vector.
c. Clustering refers to the task of grouping unlabelled objects so that data in
the same group are similar to each other than to those in other clusters. In
the figure there are two clusters identified by green and yellow ovals that
naturally group the data (red circles). d. Semi-supervised classification is similar
to regular classification but some data points do not have a label (white circles) and their label
has to be inferred from the data.
} \label{fig:supervised}
\end{figure}

Machine learning problems where the training data encompasses input vectors
paired with corresponding target output vectors, where target conveys that such a 
vector corresponds to the ideal output given the input vector, are known as supervised 
learning tasks~\cite{Bishop2006}. Starting with training dataset, the learning algorithm infers a function to make
predictions about the ideal output values for unseen input vectors. The system is thus able
to infer output vectors for any new input after sufficient training. Examples in this
category include classification, where the aim is to assign
each input vector to one of a finite number of discrete categories, and the task of
regression, where the desired output is a vector with continuous variables.
Classification is useful, e.g., for the problem of recognizing images of handwritten
digits and assigning them the most likely digit the images represent. Regression
can be used to deal with the problem of determining the orbits of bodies around
the sun from astronomical data and to extrapolate the value of observables from
simulations on finite systems to the thermodynamic limit~\cite{sandvik2010}.
Both classification and regression are illustrated in Fig.~\ref{fig:supervised}a and Fig.~\ref{fig:supervised}b,
respectively.

\subsection{Unsupervised learning}
A wide range of tasks in machine learning include situations where the training data
is composed of a set of input vectors without a corresponding target output~\cite{Bishop2006}.
Unsupervised learning studies algorithms and their ability to infer functions to discover hidden
structures in the data~\cite{Bishop2006}. Examples of tasks in unsupervised learning problems include
the discovery of groups of similar examples within the data, a task known as clustering (illustrated in Fig.~\ref{fig:supervised}c),
as well as density estimation, where the objective is to estimate the underlying
probability distribution associated with the data. Another useful unsupervised learning
task is in the low-dimensional visualization of high-dimensional data in two or three
dimensions while retaining the spatial characteristics of the original data as much as possible.

\subsection{Semi-supervised learning}
Semi-supervised learning falls between supervised and unsupervised learning~\cite{10.5555/1841234}.
It refers to a machine learning approach where a small amount of labelled data is combined with a large
amount of unlabelled data during training for classification, as illustrated in Fig.~\ref{fig:supervised}d.
Semi-supervised learning is thus extremely useful in research areas where the acquisition of labelled
data is expensive, e.g., when the collection of labelled data requires a skilled human (e.g. a
professional translator) a physical experiment or an  expensive numerical simulation. The costs
associated with the labelling process can impede the development of large labelled training data,
while acquisition of unlabelled data is relatively inexpensive in some settings. Semi-supervised learning
has been  applied to the image captioning problem and video transcription. In physics
it has been applied to the classification of phases of matter from snapshots of Monte Carlo
simulations without labelled data, as well as to the problem of efficiently
sampling rare trajectories of stochastic systems~\cite{oakes2020}.

\subsection{Reinforcement learning}
Reinforcement learning develops algorithms concerned with the problem of discovering
a set of actions that maximize a numerical reward signal.\cite{10.5555/551283} The
learning algorithm is never directly exposed to examples of optimal actions to take,
it must instead discover them by a process similar to trial and error.  In many cases,
actions affect not only the immediate reward but also subsequent actions and rewards. 
In some cases, the reward signal may come only after having executed multiple actions, which
may be discrete or continuous and can be high dimensional. An central concept in the formulation of 
a reinforcement learning algorithm is the policy, which defines the sets of actions the agent should perform
at a given time. It is a function that maps the state of an environment where the agent acts 
to concrete actions to be taken with the objective of maximizing a specific reward.    
Reinforcement learning augmented by deep learning has successfully learned
policies from high-dimensional sensory input for game playing achieving human-level
performance in several challenging games including Atari 2600~\cite{mnih2015} as well as the board game
Go~\cite{silver2017}. The success of reinforcement learning has been also applied to the 
control of quantum systems  as well as to the optimization of  quantum error correction codes, 
one of the key ingredients for fault-tolerant quantum computation.

\section{Machine learning in simulations of strongly correlated fermions}
Some of the earliest applications of machine learning techniques to quantum matter arose
in the context of molecular systems. Here, machine learning has been used to 
accurately model atomization energies~\cite{PhysRevLett.108.058301} based on a dataset 
from energies computed with with hybrid density-functional theory. This work demonstrated 
the potential applicability of supervised machine learning algorithms for the acceleration 
and prediction of atomization energies across the molecular space and inspired a wide 
variety of machine learning applications in molecular and materials science~\cite{Butler2018}. 
In the context of density functional theory (DFT), a proof-of-principle demonstration based 
on a system of free fermions showed that density functionals can be
accurately  approximated using kernel ridge regression~\cite{PhysRevLett.108.253002}.
These works motivated a wide array of machine learning applications of data-enabled chemistry and density
functional theory~\cite{doi:10.1002/qua.25040,PhysRevMaterials.3.063802} with the aim of predicting,
accelerating~\cite{pilania2013,schmidt2019,bartok2017} and improving the prediction of
atomic-scale properties of materials and chemical systems (including uncertainty
estimation~\cite{musil2019}) reaching quantum chemical accuracy using DFT~\cite{bogojeski2019}.
Proposals to bypass the solution of the Kohn-Sham equations in the DFT have demonstrated 
acceleration of simulations of materials and molecules
~\cite{Chandrasekaran2019,PhysRevA.100.022512,Brockherde}. These approaches reduce the
computational cost of DFT to linear in the size of the systems, making it orders
of magnitude faster, while providing a high-fidelity emulation of exact Kohn-Sham DFT.

Another pioneering machine learning application to strongly correlated quantum many-body quantum
systems arose in the context of dynamical mean-field theory~\cite{RevModPhys.68.13}, where
the authors in Ref.~\cite{PhysRevB.90.155136} successfully applied kernel
methods~\cite{hastie_09_elements-of.statistical-learning} to find the Green's
function of the Anderson impurity model. 

\section{Machine learning phases of matter in synthetic and experimental data}

Machine learning has been actively applied to problems in classical and quantum physics and
promises to become a basic research tool with the potential for scientific discovery in
the study of strongly correlated systems. Early ideas related to the application of machine
learning to condensed matter systems arose nearly a decade ago in the context of glasses and other
complex amorphous physical systems~\cite{ronhovde2011,ronhovde2012,nussinov2016}. The main objective 
of this research line is the automatic characterization of natural hidden structures in amorphous materials, which
are notoriously difficult to analyze theoretically and experimentally, through a multi-scale network 
based approach for the data mining of such structures. One of the most elementary machine 
learning ideas applied to physical systems is in the classification of phases of matter, either in 
synthetically generated, experimental data or a combination of simulated and experimental data. In particular
Ref.~\cite{carrasquillaMachineLearningPhases2017} demonstrated that neural network technology,
can be used to encode and discriminate phases of matter and phase transitions in classical
and quantum many-body systems, including the determination of critical exponents. Beyond
classification of simulated data, Ref.~\cite{carrasquillaMachineLearningPhases2017}
shows that convolutional neural networks can represent ground states of quantum many-body
systems, specifically the ground state of the toric code~\cite{kitaev2003a}.  This idea was 
extended to the realm of  quantum physics in the context of quantum Monte Carlo simulations 
of strongly correlated fermions in Ref.~\cite{PhysRevX.7.031038,broecker2017b}.  The key message in
Ref.~\cite{PhysRevX.7.031038,broecker2017b} is that machine learning techniques offer an alternative 
approach toward distinguishing ground state and finite-temperature phases of the strongly correlated 
fermions from numerically simulated configurations based on quantum Monte Carlo, even in the presence 
of a sign problem~\cite{avella2012}. A wide variety of data-driven machine learning techniques have 
been applied to simulations of classical and quantum systems based upon innovative supervised, 
unsupervised, and semi-supervised learning techniques to discover and analyze classical and 
quantum phase transitions in equilibrium~\cite{nieuwenburg2017,PhysRevB.94.195105,
PhysRevE.95.062122,PhysRevB.96.184410,PhysRevE.97.013306,PhysRevB.97.134109,PhysRevX.7.031038,broecker2017,
PhysRevE.96.022140,doi:10.7566/JPSJ.86.063001,PhysRevB.96.245119,PhysRevB.97.045207,
PhysRevB.96.205146,PhysRevB.96.144432,broecker2017a,PhysRevB.97.205110,PhysRevB.97.174435,
PhysRevB.96.195138,PhysRevB.98.174411,PhysRevB.99.060404,PhysRevB.99.104410,PhysRevE.99.062107,lozano-gomez2020,kottmann2020,salcedo-gallo2020}
including phases of matter characterized by topological order~\cite{carrasquillaMachineLearningPhases2017,PhysRevB.97.115453,rodriguez-nieva2019,holanda2019,greplova2020}
and out-of-equilibrium systems~\cite{PhysRevLett.121.245701,PhysRevB.97.134109,PhysRevLett.120.257204,
PhysRevB.98.060301,PhysRevLett.121.255702,PhysRevE.99.023304,PhysRevB.100.075102}. 
Unsupervised learning has also been used to model the thermodynamic observables for physical systems in thermal
equilibrium~\cite{PhysRevB.94.165134,10.5555/3122009.3242020,PhysRevB.99.054208,PhysRevE.100.052312}. Ref.~\cite{pang2018} 
uses supervised learning with a deep convolutional neural network to identify the equation of state in 
the relativistic hydrodynamic simulations of heavy ion collisions, which is one important goal of high-energy 
heavy-ion experiments. Additionally, machine learning techniques have been applied to the discovery of physical concepts and 
effective models~\cite{PhysRevB.101.241105}
from data and to the identification of physical theories~\cite{PhysRevE.100.033311,PhysRevLett.124.010508,wang2019},
the discovery of symmetry and conserved quantities~\cite{wetzel2020}, and even to generate 
computer-inspired scientific ideas~\cite{Krenn1910}.  

While results related to the analysis and processing of simulation data has provided
significant insight into the potential for machine learning techniques to impact scientific discovery,
one of the most important application where machine learning can truly excel is in the analysis
of complex experimental data, which is often plagued with noise and other imperfections.
In this context, machine learning techniques have been applied to the analysis of
data coming from ultracold atom experiments taken with a quantum gas microscope~\cite{bakr2009},
demonstrating that machine learning has the capability to distill microscopic mechanisms and hidden
order in experimental data~\cite{bohrdt2019,khatami2020,casert2020}. In particular,  Ref.~\cite{bohrdt2019} found
evidence that machine learning applied experimental snapshots from quantum many-body states can
help distill the most predictive theory among a multitude of competing theories.
Additionally, artificial neural networks and deep-learning techniques have been used
to identify quantum phase transitions from single-shot experimental momentum-space density
images of ultracold quantum gases displaying results that go beyond conventional
methods in terms of accuracy in the detection of phase transitions~\cite{rem2019}.
Another notable application of machine learning to experimental data is in the analysis of 
complex electronic-structure images~\cite{zhang2019}. Ref.~\cite{zhang2019} reports
the development and training of a set of artificial neural networks aimed at
recognizing different types of order hidden in electronic quantum matter images
from carrier-doped copper oxide Mott insulators. Here, the neural networks discovered
the existence of a lattice-commensurate, four-unit-cell periodic,
translational-symmetry-breaking state.  In a similar vein, Ref.~\cite{ziatdinov2016}
demonstrated that statistical learning applied to scanning tunnelling microscopy
data can be used to uncover relevant electronic correlations in the gold-doped
BaFe$_2$As$_2$. Ref.~\cite{samarakoon2019} uses an autoencoder to extract model Hamiltonians 
from experimental data, and to identify different magnetic regimes in the system. The system
considered in Ref.~\cite{samarakoon2019} is the spin ice material Dy$_2$Ti$_2$O$_7$, for which 
the authors find an optimal Hamiltonian capable of accurately predicting the temperature and field 
dependence of both magnetic structure and magnetization, among other properties. 
The observations in Ref.~\cite{ziatdinov2016,zhang2019,samarakoon2019} offer a glimpse
into the potential for discovery that machine learning techniques such as k-means, principal 
component analysis~\cite{Bishop2006}, neural networks, autoencoders, and other statistical learning 
techniques~\cite{10.1109/TSP.2009.2025797} can offer as a complementary view on the physical 
nature of complex phases of matter beyond traditional methods of analysis of experimental data.

\section{Renormalization group and its relation to machine learning}

The renormalization group (RG) refers to the mathematical infrastructure that
facilitates the study of the changes of a physical system when viewed at different
length scales and is currently the theoretical bedrock for our understanding of phase
transitions and critical phenomena~\cite{RevModPhys.70.653}. RG methods remain also
key to our understanding of modern condensed-matter theory and particle physics.

Motivated by the structural and inner workings of deep neural networks, which
have been seen to operate by extracting a hierarchy of increasingly higher-level
concepts in its layers, a number of studies have established a series of connections
between RG and deep learning~\cite{beny2013,mehta2014,koch-janusz2018,PhysRevE.97.053304,li2018a,lenggenhager2019,koch2019,chung2019}
One of the earliest connections between RG and deep learning appeared in Ref.~\cite{beny2013}.
The author compares RG and deep learning and establishes that a foundational tensor network
architecture, i.e., the multiscale entanglement renormalization ansatz (MERA)~\cite{PhysRevLett.101.110501},
can be converted into a learning algorithm based on a generative hierarchical Bayesian network model.

A very influential paper about the connection between RG and deep learning~\cite{mehta2014} describes
an exact mapping between the variational RG and a specific deep learning architecture
based on stacked Restricted Boltzmann Machines (RBMs)~\cite{hinton2012,melko2019}. In addition
to the exact mapping, the authors in Ref.~\cite{mehta2014} numerically explore the training of their
deep architecture using data from the two-dimensional Ising model and conclude that these models
are trained to approximately implement a coarse-graining procedure similar to Kadanoff's block
renormalization. 

On a more practical level, Ref.~\cite{koch-janusz2018} proposed an RG scheme
based on mutual information maximization that automatically identifies the relevant degrees
of freedom for the RG decimation procedure without any prior knowledge about the system. The authors apply their
techniques to the two-dimensional Ising model and a dimer model. For the dimer model, the algorithm
subtly discovers the adequate degrees of freedom for the RG procedure; in contrast to the Ising case where
the degrees of freedom are groups of spins,the correct degrees of freedom to perform RG on are not dimers,
but rather effective local electric fields. Ref.~\cite{li2018a} introduces a variational renormalization
group approach based on a reversible generative model with hierarchical architecture and apply it
in the identification of mutually independent collective variables of the Ising model as well as to the
acceleration of Monte Carlo sampling. Such collective variables play an important role in the acceleration of
molecular simulations based on metadynamics~\cite{barducci2011}, and thus these machine-learning inspired 
RG techniques may potentially impact the important area of molecular dynamics.

\section{Neural-network quantum states and their applications}
A central concept common to problems in simulations of quantum matter and quantum technologies
is the many-body wavefunction, which is one of the most complex mathematical objects in physics. Remarkably,
the power of important quantum technologies relies on our ability to accurately control, measure, and
characterize the wavefunction of large but brittle quantum systems. As a consequence, an accurate specification
and control of the state of a quantum system remains a key research topic in several quantum technology
labs around the world. In a series of developments, researchers have taken the fundamental
viewpoint that the state of a quantum system, whose exponential complexity is reminiscent of the
``curse of dimensionality'' encountered in machine learning, can be understood as a generative model
of phenomena at the microscopic scale. This has led to the idea of neural-network quantum states, which
are representations of a quantum state based on powerful function approximators for the wavefunction based
upon neural networks.

The idea of using neural networks to study quantum systems has a relatively long history.
Some of the early applications of neural networks in quantum physics appeared in the early
90's, surprisingly during the artificial intelligence winter~\cite{darsey1991,androsiuk1993}.
These studies used regression and data from exact solutions to obtain accurate potential
energy surfaces of a two-dimensional harmonic oscillators. Despite their simplicity,
it was already clear that direct application of neural networks could be used to investigate
questions related to basic issues of physics and chemistry.

\begin{figure}
\centering
{%
\resizebox*{7cm}{!}{\includegraphics{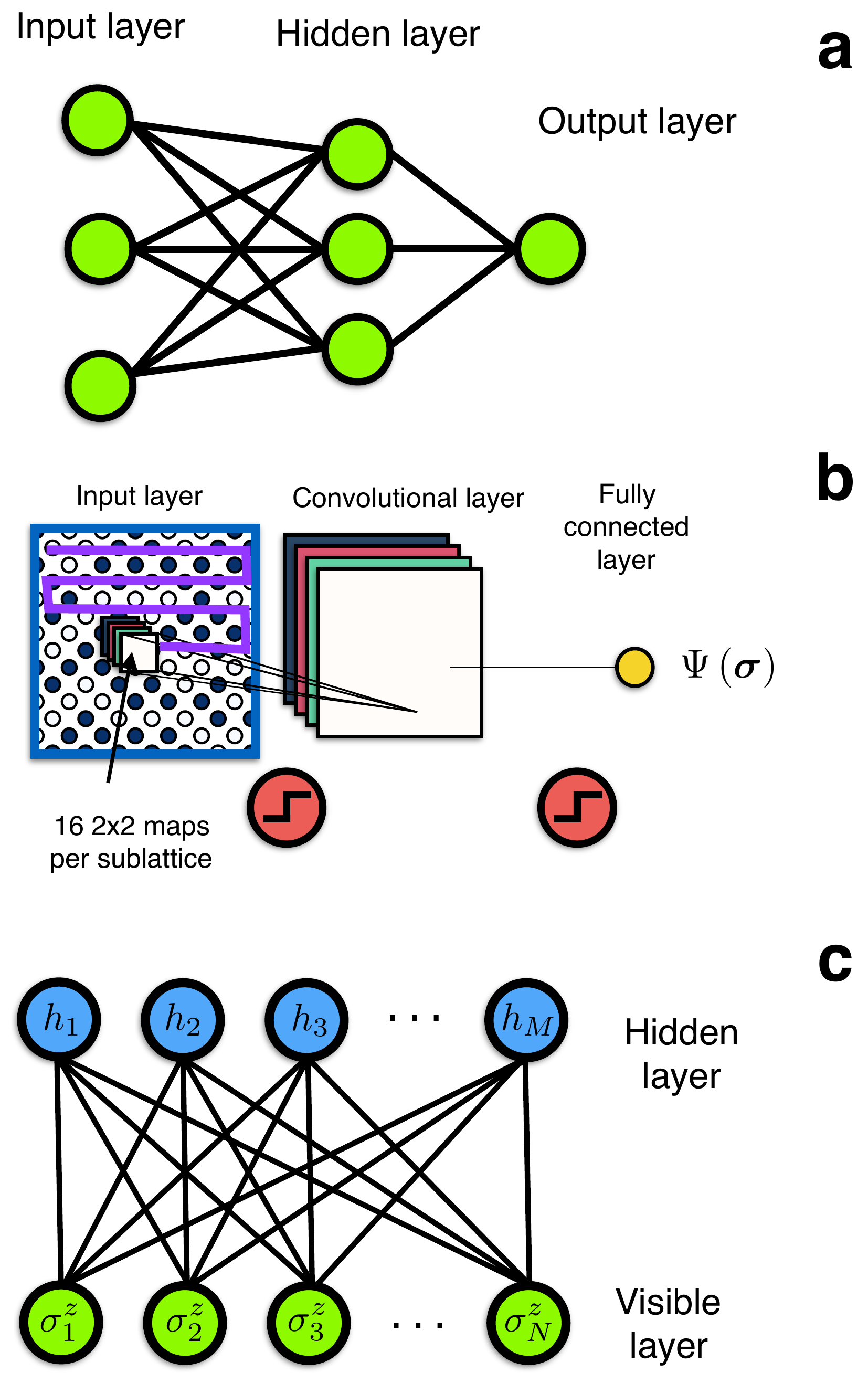}}}\hspace{5pt}
\caption{Three examples of neural network architectures to build quantum states. a. A fully connected
neural network with 3 layers (input, hidden, and output layer from left to right). In a fully connected
neural network every node in a layer forms fully-connected network with those of the adjacent
layers but not with those within the same layer. Usually, in  a neural network quantum state the
input layer corresponds a spin/electron configuration and the output layer determines the amplitude of the
wavefunction for the given spin/electron configuration. b. A convolutional neural network with one convolutional
layer fully connected to the output layer and perceptron activations. The input corresponds to a spin configuration
$\boldsymbol{\sigma}$ (represented as a two-dimensional array of binary values) and the output quantifies the
amplitude of the ground state of the toric code $\Psi\left(\boldsymbol{\sigma}\right)$. c. An RBM with $M$ hidden neurons and a visible layer with $N$ spins as the input. For each
value of the spin configuration $\sigma_1^z,\cdot,\sigma_N^z$ the neural network returns the value of
the wavefunction $\Psi(\boldsymbol{\sigma})$.} \label{fig:NQS}
\end{figure}

Some of the earliest neural-network quantum states were introduced in the 90's a well. In
Ref.~\cite{lagaris1997} the authors used a fully connected feed-forward neural network depicted
in Fig.~\ref{fig:NQS}a to represent the wavefunction of a system of a single particle and found
accurate solutions to the Schr{\"o}dinger equation in several scenarios including muonic atoms,
the Morse potential, two-dimensional potentials, three coupled unharmonic oscillators, as well
as the Dirac equation for muonic atoms. The parameters of the neural network were optimized using
backpropagation and the steepest descent method with an objective function similar to functions
used in the local-energy method in electronic structure calculations~\cite{RevModPhys.32.313}.
Likewise, the author of Ref~\cite{sugawara2001} used a feed-forward neural network to represent
the wavefunction. The parameters of the network were optimized using a micro-genetic algorithm
so that the neural network satisfied the Schr{\"o}dinger equation for a one-dimensional harmonic
oscillator.

The recent resurgence of interest in machine learning in the physical sciences has motivated a new
playground for variational calculations and exact representations of quantum states based on neural
networks~\cite{carleo2017,melko2019}. One of these early examples was developed in
Ref.\cite{carrasquillaMachineLearningPhases2017}, where ground states of the toric
code~\cite{kitaev2003a} were expressed in terms of a convolutional neural network with a
structure depicted in Fig.~\ref{fig:NQS}b. Here, the idea is to impose the constraints induced
by the toric code Hamiltonian $H_{\text{toric}}=-J_p\sum_{p}\prod_{i\in p}\sigma_i^z-J_v \sum_{v}\prod_{i\in v}\sigma_i^x$,
in particular those imposed by the $J_p$ term, directly in a convolutional neural network. This
solution takes inspiration from the construction of the ground state of the toric code in terms
of projected entangled pair states in that local tensors project out states containing plaquettes with
odd parity~\cite{PhysRevLett.109.260401}. The convolutional layer contains 16 $2\times2$ filters per
sublattice with unit stride in both directions and periodic boundary conditions. The outcome of the
convolutional layer is fully connected to a perceptron neuron in the output layer which represents
the wavefunction in the computational basis $\Psi(\boldsymbol{\sigma})$. Here 
$\boldsymbol{\sigma}=(\sigma^{z}_1,\sigma^{z}_2,\cdots,\sigma^{z}_N)$ represents a spin-$1/2$
configuration of the computational basis for $N$ spins.

The most influential study on neural-network quantum states introduced a family of variational
wavefunctions based upon a restricted Boltzmann machine (RBM) in Ref.~\cite{carleo2017}
Originally invented by Paul Smolensky~\cite{Smolensky1986} and popularized by Geoffrey
Hinton~\cite{doi:10.1162/089976602760128018}, this architecture has been recently repurposed as
a representation of quantum states~\cite{carleo2017,melko2019}. In this context, the RBM  has been
used to approximate the ground state of prototypical systems in condensed matter physics such
as the transverse field Ising and Heisenberg models in one and two dimensions~\cite{carleo2017,PhysRevB.100.125131,PhysRevB.96.205152},
the Hubbard model~\cite{PhysRevB.96.205152}, models of frustrated magnetism~\cite{PhysRevB.100.125131},
the Bose-Hubbard model~\cite{mcbrian2019a,vargas-calderon2020}, ground states of molecules~\cite{choo2019a}, to model
spectral properties of many-body systems~\cite{PhysRevLett.121.167204},  as well as to study
non-equilibrium properties of quantum systems~\cite{carleo2017,PhysRevB.98.024311}. While the 
simulation of real-time dynamics remains a challenge, Ref.~\cite{schmitt2019} has demonstrated 
that neural networks represent a promising tool for the simulation
of dynamical systems with simulations attaining time scales that compare well or even exceed simulations
based on tensor networks. The RBM has also been applied to the study of many-body open quantum 
systems~\cite{PhysRevB.99.214306,PhysRevLett.122.250502,PhysRevLett.122.250501,PhysRevLett.122.250503}
as well as a tool to perform approximate quantum state tomography for many-body
systems~\cite{torlai2018,PhysRevLett.120.240503,carrasquilla2019}. A complementary study to the neural-based
tomographic schemes is presented in Ref.\cite{huang2020}, where the authors present an efficient 
method for constructing an approximate classical description of a quantum state using very few 
measurements of the state with well established performance guarantees.  

Additionally, since the RBMs are amenable to analytical treatment, they have been used
to find exact representations of quantum states of a wide array of quantum many-body systems.
This includes ground states of one-dimensional symmetry-protected topological cluster state~\cite{PhysRevB.96.195145} and the
toric code states~\cite{PhysRevB.96.195145,PhysRevB.97.085104} as well as other topologically
ordered states of matter such as the ground states of double semion and twisted quantum double
models, states of the  Affleck, Lieb, Kennedy and Tasaki (AKLT) model and two-dimensional CZX model,
states of stabilizer Fracton models with fracton topological order~\cite{PhysRevB.99.155136},
 fractional quantum Hall state~\cite{PhysRevX.8.011006}, among others~\cite{PhysRevB.99.155136,PhysRevX.8.011006}.

The RBM has been characterized theoretically as well, including their representational power as
classical probability~\cite{leroux2008} and as a quantum state~\cite{gao2017}. Their entanglement
properties and capacity have been studied in Ref~\cite{PhysRevX.7.021021,PhysRevB.97.085104},
as well as their relation with tensor networks states has been carefully established in
Ref.~\cite{PhysRevB.97.085104,Clark_2018,PhysRevX.8.011006}. The possibility to directly impose 
symmetries on the RBM has also been studied~\cite{carleo2017}, including non-Abelian or anyonic 
symmetries~\cite{PhysRevLett.124.097201}

In spite of the existence of heuristic algorithms to sample and compute their normalization constants,
we would like to highlight  that the sampling of families of neural network quantum states, in particular
the RBM family, is a computationally intractable task~\cite{10.5555/3104322.3104412}. To alleviate these issues,
RBMs can be physically implemented by neuromorphic hardware which has the potential to enable better sampling.
Ref.~\cite{PhysRevB.100.195120} leverages this capability to speed up sampling RBMs and finds that the use 
of neuromorphic chips has great potential to reduce computation times of calculations based on RBMs, thereby 
extending the range of tractable system sizes represented by these neural architectures.

Going beyond RBMs, quantum states based on convolutional neural networks have been shown to
accurately model ground states of complex frustrated spin systems~\cite{PhysRevB.98.104426,PhysRevB.100.125124,szabo2020}
as well as finite-temperature states~\cite{irikura2019} and bosons on the lattice~\cite{doi:10.7566/JPSJ.87.014001}.
Wavefunctions and other neural network representation of the quantum state based on autoregressive models
allow for uncorrelated sampling from the wavefunction, unlike traditional variational Monte Carlo methods~\cite{becca2017}, 
where an expensive Markov chain introduces potential biases in the calculations of observables and during the optimization
of the quantum state. Examples of this include a recurrent neural network representation of the quantum state
based on generalized measurements~\cite{carrasquilla2019}, as well as neural autoregressive quantum states~\cite{PhysRevLett.124.020503}
and a recurrent neural network wavefunctions~\cite{hibat-allah2020,roth2020}, both of which produce state-of-the-art
approximations to the ground states of prototypical models in condensed matter physics.

When considering wavefunction approximations for fermionic systems such as molecules, real materials, as
well as fermionic models on the lattice such as the Hubbard model, it is fundamental to account for the
essential anti-symmetry of the wavefunction ansatz. This can be achieved in several different ways. The
simplest approach relies on the specification of variational fermion wavefunctions based on Slater
determinants arising from a mean-field Hamiltonian that best match the interacting ground state.
Naturally, this approach misses some quantum fluctuations which can be reintroduced through a two-body
Jastrow factor supplementing the mean-field treatment, often called Slater-Jastrow wavefunctions. The
simplest neural extension of the Slater-Jastrow wavefunction consists in replacing the Jastrow factor
with a neural network. As shown in Ref.~\cite{PhysRevB.96.205152}, a mean-field ansatz is supplemented 
with an RBM leading to a substantial improvement of the accuracy beyond that achieved 
by each method separately for in the Heisenberg and Hubbard models on square lattices.

Originally introduced by Feynmann and Cohen~\cite{PhysRev.102.1189}, the backflow adds correlation to
a mean-field ground state by transforming the single-particle orbitals in a configuration-dependent
way. Luo and Clark~\cite{PhysRevLett.122.226401} introduced a neural network backflow, where
a neural network dresses a mean-field state endowed with the Fermi-Dirac statistics which enables a
systematic improvement over mean-field states and provides excellent results on the two-dimensional
Hubbard model. Similar ideas to the original backflow idea have also been recently applied to a
variety of atoms and small molecule systems directly continuum~\cite{hermann2019,pfau2019}. These 
Fermionic neural networks predict the dissociation curves of the simple molecules and the hydrogen chain,
 to significantly higher accuracy than the coupled cluster method in chemistry~\cite{pfau2019}. Finally,
yet another strategy to impose Fermi-Dirac statistics is by transforming the original fermionic
into a spin system through a Jordan-Wigner transformation~\cite{jordan1928}. After transforming the
original Hamiltonian to a bosonic one, then one can use neural-network quantum states to perform
electronic structure calculations, as demonstrated in Ref.~\cite{choo2019a}.

To conclude, we mention a recent review of neural-network quantum states~\cite{jia2019} which
illustrates various representations for pure and mixed quantum states and discusses their
physical properties and recent progress related to the application of neural-network quantum
states to tomography and the simulation of many-body quantum systems~\cite{jia2019}.

\section{Machine learning acceleration of Monte Carlo simulations}
Quantum Monte Carlo (QMC) refers to a wide array of computational methods based on Monte
Carlo techniques with the aim of studying of quantum many-body systems, including many
methodologies to determine ground-state, excited-state or finite-temperature equilibrium as
well as non-equilibrium properties of a variety of quantum systems. Although Fermi's famous
suggestion of the first QMC algorithm was already acknowledged in a 1949 paper by Metropolis 
and Ulam~\cite{metropolis1949}, QMC methods remain a powerful and broadly applicable computational 
tool for finding accurate solutions of Schr{\"o}dinger equation for atoms, molecules, quantum 
spin systems, and materials.

While Monte Carlo (MC) simulations remain a powerful tool  in the study of classical and
quantum many-body systems, a key issue with MC is the lack of a general and efficient
update algorithms for large size systems close to critical points and other challenging
statistical physics problems, where MC algorithms suffer a slow convergence~\cite{Newman1999}.
Inspired by recent advances in machine learning, Ref.~\cite{PhysRevB.95.041101,PhysRevB.95.035105,PhysRevE.96.051301}
simultaneously proposed general-purpose methods to accelerate MC simulations.
Ref.~\cite{PhysRevB.95.041101} introduced a method dubbed self-learning Monte Carlo (SLMC),
in which an efficient update algorithm is first learned from the training configurations generated
in trial unaccelerated simulation and then used to speed up the simulation. The
authors demonstrate their technique on a spin model at the phase transition point, achieving a
10 to 20 times speedup. Simultaneously, Ref~\cite{PhysRevB.95.035105} introduced a general
strategy to try to overcome the MC slow down by fitting the unnormalized probability of the
physical model to a feed-forward neural network. The authors then utilize the neural network
for efficient MC updates and to speed up the simulation of the original physical system. This
technique was applied to the  Falicov-Kimball model~\cite{PhysRevLett.22.997} where
improved acceptance ratio and autocorrelation times near the phase transition point
were observed~\cite{PhysRevB.95.035105}. 

The SLMC scheme has been extended in several directions including
its application to interacting fermion models within the framework of determinant QMC~\cite{PhysRevB.96.041119}. 
The SLMC has been augmented with deep neural networks
and applied to the reduction of complexity of simulating quantum impurity models~\cite{PhysRevB.97.205140}. Likewise,
Ref.~\cite{PhysRevB.101.241105} proposes a machine learning approach that optimizes an effective model based
on an estimation of its partition function and applies the method to the single impurity Anderson model and double quantum dots.
Ref.~\cite{PhysRevB.98.041102,PhysRevB.100.020302} improved the efficiency of SLMC applied
to the Holstein model, which represents one of the most fundamental many-body descriptions of electron-phonon coupling.
The authors in Ref.~\cite{PhysRevB.98.041102} endow the effective description of the action of
the model with physical information by defining an effective bosonic Hamiltonian for the phonon
fields which incorporates a global $Z_2$ symmetry of the original Holstein model. This  leads
to an outstanding reduction of computational complexity from $O(L^{11})$ to $O(L^7)$, enabling
the evaluation of the metal-to-charge density wave transition temperature to an order of magnitude
higher accuracy than previously available~\cite{PhysRevB.98.041102}. Additionally, the SLMC has
been extended to the framework of a continuous-time MC method with an auxiliary field for
quantum impurity models~\cite{PhysRevB.96.161102}, as well as to the framework of Hybrid MC
in the context of first-principles molecular dynamics based on density functional theory~\cite{nagai2019,nagai2018}.
SLMC has also been applied to the study of the Gross-Neveu-Yukawa chiral-Ising quantum critical point
with critical bosonic modes coupled with Dirac fermions~\cite{PhysRevB.101.064308}, where
the authors obtain challenging quantities including a comprehensive set of critical exponents
as well as the conductivity of the Dirac fermions of the theory.

The continuous time QMC method remains one of the best impurity solver
for dynamical mean-field theory (DMFT). Such an impurity problem describes how the electrons on an
impurity site interact with electrons in a bath and whose solution is the most challenging
and computationally expensive part of the DMFT approach. Ref.~\cite{PhysRevB.100.045153}
utilizes a machine learning technique, specifically a convolutional autoencoder~\cite{Goodfellow-et-al-2016},
to reduce the computational complexity of the DMFT procedure. While the machine learning approach is not exact,
the authors demonstrate that their approach retains the accuracy in the estimation of
important correlation functions, as demonstrated through careful comparisons to the exact
solution of the impurity solver. 

Projector quantum Monte Carlo (PQMC) techniques are powerful computational methods to
simulate properties of quantum many-body systems~\cite{becca2017}. The success of these
methods crucially rely on the our ability to construct an accurate guiding wavefunction, which in the
standard formulation, is optimized in a separate simulation using variational Monte Carlo~\cite{becca2017}. 
Ref.~\cite{PhysRevB.98.235145} investigates the use of variational wavefunctions based upon 
unrestricted Boltzmann machines~\cite{Goodfellow-et-al-2016} as guiding functions in 
PQMC simulations of quantum spin models. The authors
in Ref.~\cite{PhysRevB.98.235145} demonstrate that the optimized unrestricted neural-network 
states as guiding function leads to an increased efficiency of the PQMC algorithms, drastically 
reducing the most relevant systematic bias of the algorithm, i.e., the bias due to the finite 
random-walker population~\cite{becca2017}. In a similar vein, Pilati, Inack and Pieri propose 
another class of SLMC which augments the PQMC with a neural network~\cite{PhysRevE.100.043301}. 
Here, the authors of Ref.~\cite{PhysRevE.100.043301} develop PQMC simulations guided by an adaptive 
RBM wavefunction that is optimized along the PQMC simulation via unsupervised machine learning, 
avoiding the need of a separate variational optimization. Thus, beyond demonstrating an excellent 
convergence of the PQMC procedure,  this technique provides an accurate ansatz for the ground-state 
wavefunction, which is obtained by minimizing the Kullback-Leibler divergence~\cite{Goodfellow-et-al-2016} 
with respect to the PQMC samples. 

Deep reinforcement learning has also been used in conjunction with MC 
simulations~\cite{PhysRevE.99.062106}. Zhao {\it et al} develop a deep reinforcement 
learning framework where a machine agent is trained to search for a policy to
generate  ground states for the square ice model~\cite{PhysRevE.57.1155}, which belongs
to the family of ice models used to describe the statistical properties of the hydrogen 
atoms in water ice. The authors' analysis of the learned policy  and the state value 
function reveals that the ice rule and loop-closing condition are learned without prior information
about other than the Hamiltonian of the system. Importantly, the authors envisage that it is 
possible to extend this framework to other physical models and quantum systems such as quantum
spin ice~\cite{gingras2014}, the toric code~\cite{kitaev2003a}, and other models endowed 
with physical constraints which induce long autocorrelation times in with MC simulations. Reinforcement
learning has also been used to optimized the temperature schedule of simulated annealing applied to
binary optimization problems~\cite{mills2020}. 

Neural autoregressive models have also been applied to the solution of classical
statistical mechanics in a classical variational setting~\cite{PhysRevLett.122.080602} and to 
extrapolate observables beyond the region of training~\cite{sprague2020}. The method in 
Ref.~\cite{PhysRevLett.122.080602} extends the variational mean-field approach  using 
autoregressive neural networks, computes the variational free energy, estimates physical quantities such as entropy, 
magnetization and correlations, and generates uncorrelated samples. The authors in 
Ref.~\cite{PhysRevLett.122.080602} apply their methodology to several systems including 
two-dimensional Ising models, the Hopfield model, the Sherrington-Kirkpatrick model, 
and the inverse Ising model, where they find excellent agreement between the exact solution
of the models and the neural variational approach. Similarly, Ref.~\cite{mcnaughton2020}
explore the autoregressive neural networks for the improvement of classical MC
simulations of the two-dimensional Edwards Anderson spin glass, a paradigmatic classical model
of spin-glass theory. Likewise, Ref.~\cite{PhysRevE.101.023304} explores a general framework 
for the estimation of observables with generative neural samplers focusing on autoregressive networks
and normalizing flows that provide an exact sampling probability. These examples  anticipate that 
neural autoregressive models have potential applications in important combinatorial optimization and constraint satisfaction 
problems, where finding the optimal configurations corresponds to finding ground states of 
glassy problems, and counting the number of solutions is equivalent to estimating the zero-temperature
entropy of the system.

Lattice field theory has also emerged as an area where machine learning techniques can be used to advance 
MC simulations which traditionally suffer from the critical slowdown problem~\cite{urban2019,PhysRevD.100.034515,PhysRevD.100.011501}.
Ref.~\cite{PhysRevD.100.034515} conceived a Markov chain update scheme using a machine-learned 
flow-based generative model~\cite{pmlr-v37-rezende15} and applied it to the simulation of a $\phi^4$ theory. 
The training of the model  systematically improves autocorrelation times in the Markov chain, including 
regions where standard Markov chain MC, such as Hamiltonian Monte Carlo (HMC)~\cite{brooks2011handbook}, 
exhibit critical slowing down. The authors find that their algorithm produces ensembles of configurations
that are indistinguishable from those generated using local Metropolis and HMC  for a number of physical 
observables but leave the question about scalability and wider applicability for future studies. 
Ref.~\cite{urban2019} proposes to reduce the autocorrelation times in lattice field theory simulations via
a Generative Adversarial Network (GAN)~\cite{goodfellow2014}. Ref.~\cite{urban2019} implements the GAN as 
an overrelaxation step in combination with the traditional HMC algorithm. This allows the method  to meet all 
the statistical requirements to produce correct results by using the Metropolis-Hastings accept/reject rule. 
This combination breaks the Markov chain but effectively reduces the autocorrelation time of observables 
and correctly captures the dynamics of the theory.

\begin{figure}
\centering
{%
\resizebox*{7cm}{!}{\includegraphics{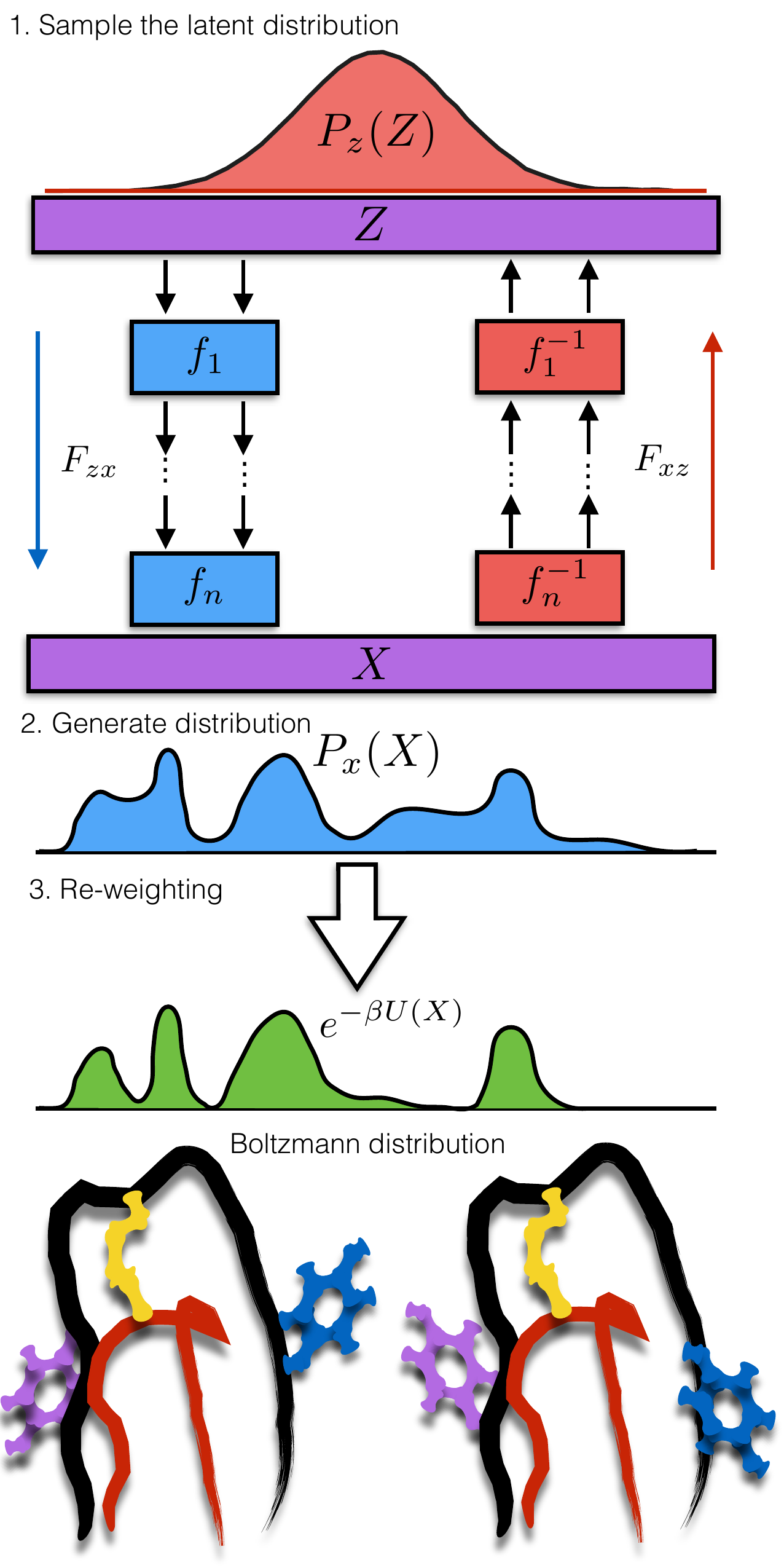}}}\hspace{5pt}
\caption{ Boltzmann generator. 1. The Boltzmann generator samples from a simple latent distribution $P_{z}(Z)$
(e.g., Gaussian distribution).  2. An invertible deep neural network (red and blue blocks) is 
trained to transform the distribution $P_{z}(Z)$ to a distribution $P_{x}(X)$ that approximates the 
desired Boltzmann distribution of a complex system $e^{-\beta U(X)}$. 3. To compute quantities of practical
interest, the samples are reweighted to the Boltzmann distribution~\cite{Newman1999}. Two configurations of the
complex system, e.g., a protein, is depicted in the figure below. 

} \label{fig:boltzmanngen}
\end{figure}

Finally, we highlight the concept of Boltzmann generators introduced in Ref.~\cite{Noeeaaw1147}. Boltzmann
generators combine deep learning and statistical mechanics to generate unbiased one-shot equilibrium 
samples of challenging condensed-matter systems and proteins described by some energy function $U(X)$ at inverse temperature
$\beta$. This is accomplished by optimizing an invertible neural network~\cite{dinh2017} to represent a 
coordinate transformation from the actual system configuration space to a modified space that is easy 
to sample. Importantly, the model is trained so that low-energy configurations in the so-called latent 
space are close to each other. Due to the invertibility of the neural model, samples in the latent 
space $Z$ can be transformed to a system configuration $X$ with approximately the right Boltzmann probability. 
This approximate sample, together with its model's likelihood $P_x(X)$, can be  
used in combination with a reweighting scheme~\cite{Newman1999} to produce correct samples from the 
Boltzmann distribution $e^{-\beta U(X)}$ for challenging systems including proteins and other difficult
condensed matter systems. The Boltzmann generator is illustrated in Fig.~\ref{fig:boltzmanngen}. 

\section{Quantum information, quantum control, and quantum computation}
\subsection{Measurements}
In this section we review recent applications of machine learning ideas to quantum 
information processing. Due to the strong connections between machine learning and quantum 
information, which stems from the fact probability and statistics lay the foundations
for these two areas of research, this field has a relatively long 
history. Thus, our aim is not to review these developments entirely, but focus on 
a set of recent and illustrative studies. 

Hentschel and Sanders propose to use machine learning technology to tackle the quantum
measurement problem, where the aim is to infer parameters of interest from measurements
on a quantum system~\cite{PhysRevLett.104.063603}. Beyond estimating the parameters, 
quantum metrology is concerned with the identification of optimal measurement strategies for 
a given parameter estimation problem. Ref.~\cite{PhysRevLett.104.063603} tackles the phase estimation 
problem through particle swarm optimization to autonomously generate adaptive feedback 
measurement policies for interferometric phase estimation problems. The setup they study 
corresponds to the estimation of an unknown phase difference $\phi$ between the two arms
of a Mach-Zehnder interferometer. The authors find that the particle swarm optimization policies 
achieve an optimal scaling of precision for single shot interferometric phase estimation.

In Ref.~\cite{greplova2017} Greplova, Andersen,  and M{\o}lmer adapt image recognition algorithms based
on  neural networks to estimate rates of coherent and incoherent processes in simulated 
quantum systems from discretized time measurement records. Their neural networks approach 
translates a quantum parameter estimation into a regression problem, which conveniently, 
does not require the characterization of quantum or classical noise. The neural network 
used for the parameter estimation comprises a one-dimensional convolutional layer endowed with
several filters followed by a pooling layer and a densely connected layer. The output layer 
provides the likely candidate values of the parameters characterizing the input signal. A measurement-related 
study is discussed in Ref.~\cite{xu2019}, where it is shown that reinforcement learning algorithms can be used 
to identify optimal quantum controls for a precise quantum parameter estimation. These studies 
showcase the power of reinforcement learning as an alternative to conventional optimal control methods.
Machine learning methods have also been used to measure logarithmic negativity from very few 
measurements~\cite{PhysRevLett.121.150503} 

\subsection{Quantum state reconstruction}
The tasks of reconstructing the quantum states and processes, which are known as quantum state 
and process tomography, respectively, are the gold standards for verification and benchmarking 
of quantum devices~\cite{cramer2010}.  While powerful, exact quantum state tomography (QST) has 
high computational complexity: the number of measurements required for an accurate reconstruction, 
the time to analyze such measurements, and the memory required to store the resulting state all scale exponentially
with the size of the system. This makes traditional QST unfeasible 
for anything except small systems. Machine learning techniques can be used to alleviate the scaling
of QST at the cost of assuming that the state under scrutiny possesses a structure amenable to a 
description using machine learning architectures~\cite{torlai2018,carrasquilla2019,PhysRevLett.120.240503}. 
Torlai {\it et al}~\cite{torlai2018} demonstrated that neural networks can be used to perform QST of 
entangled states with more than a hundred qubits. This work demonstrated that machine learning 
enables the reconstruction of traditionally challenging quantities, such as the entanglement entropy,
from experimentally accessible projective measurements. Ref.~\cite{PhysRevLett.120.240503} extended 
the approach to small mixed states and Ref.~\cite{carrasquilla2019} provided a scalable
way to reconstruct mixed states and introduced a built-in approximate certificate of the 
reconstruction which makes no assumptions about the purity of the state under scrutiny. The strategy
in Ref.~\cite{carrasquilla2019} can handle complex systems including prototypical states in 
quantum information, as well as ground states of local spin Hamiltonians. The problem of analysis
speed was addressed in Ref.~\cite{quek2018}, where a machine learning based algorithm for QST 
with adaptive measurements was seen to provide orders of magnitude faster processing while 
retaining a high reconstruction accuracy. Ref.~\cite{lohani2020} constructed a neural-network based 
QST framework from a set of coincidence measurements. The authors consider
the situation where a number of the projective measurements are not performed, which corresponds to 
the task of reconstructing a density matrix from informationally incomplete projective data. The authors find a dramatic 
improvement in the average reconstruction fidelity even when only a small fraction  of the total measurements
are performed, which suggests that the power and generalization ability of neural networks aid the
reconstruction in the absence of an informationally complete measurement.

\begin{figure}
\centering
{%
\resizebox*{7cm}{!}{\includegraphics{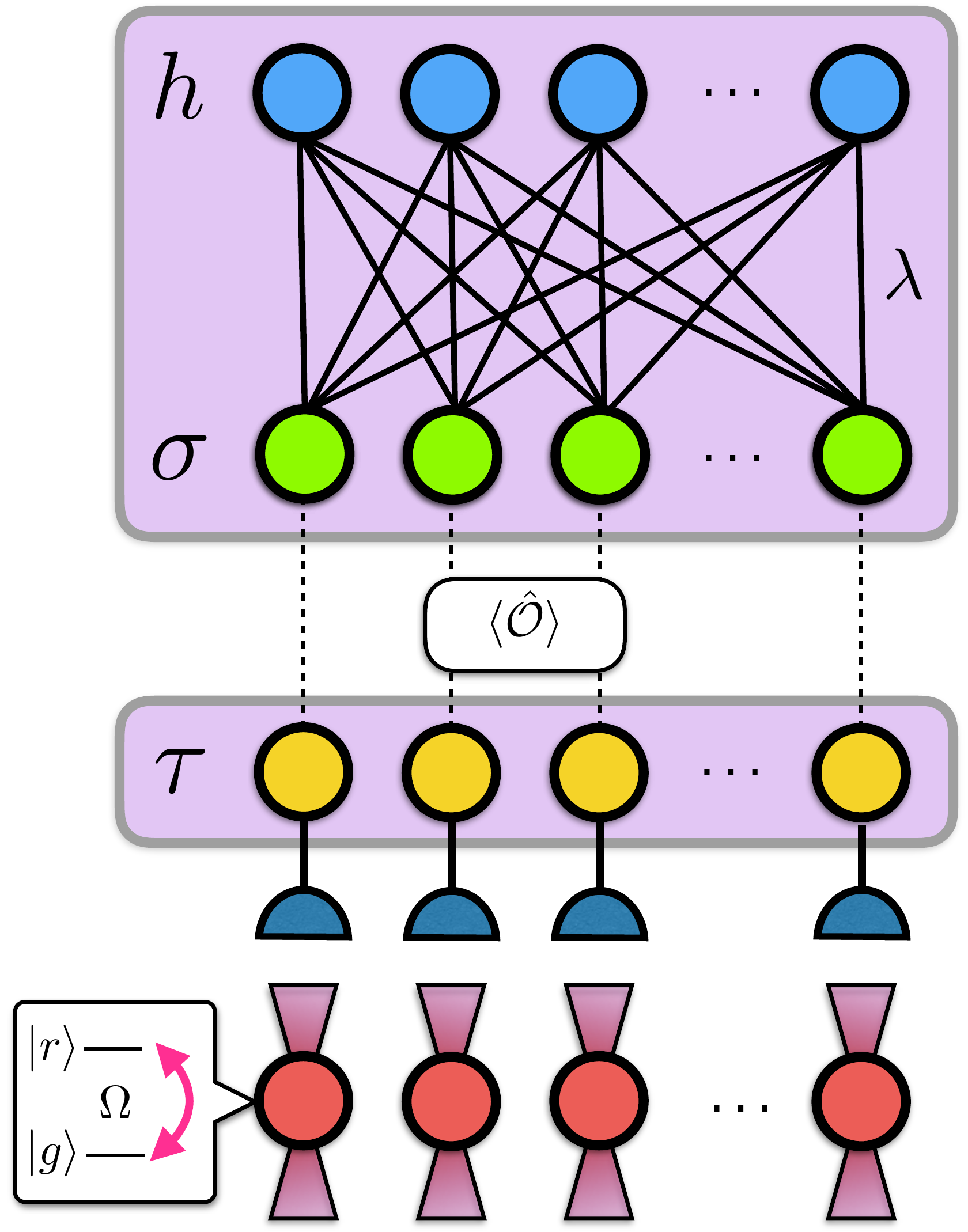}}}\hspace{5pt}
\caption{ The model of the reconstruction process in Ref.~\cite{PhysRevLett.123.230504}. 
Individual $^{87}$Rb atoms (depicted as red circles) are
trapped in an array of optical tweezers (depicted as up/down triangles behind the red atoms) 
and coupled to a Rydberg state with Rabi frequency $\Omega$. Fluorescence imaging
provides noisy measurements in the $\sigma_z$ basis. The RBM  
(blue, hidden variables $h$; green, visible spins $\sigma$) represents the reconstructed 
quantum state via a set of parameters $\lambda$. The binary data $\tau$ accessible to 
the experimentalist are included as a noise layer (yellow neurons). 
Training on this data, the RBM learns a representation of the experimental quantum state, 
which can be used to evaluate observables $\langle  \mathcal{\hat{O}} \rangle$ and R{\`e}nyi entropies. 
} \label{fig:rydberg}
\end{figure}

Most of the studies discussed so far have demonstrated the utility and scalability of machine learning approaches
to quantum state reconstruction by providing proof-of-principle demonstrations based on numerically
generated data. Here, we highlight Ref.~\cite{PhysRevLett.123.230504} which demonstrated 
quantum many-body state reconstruction from experimental data generated by a programmable 
quantum simulator based upon Rydberg atoms. The experiment, which uses 8 and 9 
atoms and has access to only a single measurement basis, applies a novel regularization technique to 
mitigate the effects of measurement errors in the training data. By exploiting structural information
about the state produced by the experiment, the quantum state reconstructions enable the 
inference of one- and two-body observables not directly accessible to experimentalists, as well as 
observables such as the R{\`e}nyi mutual information. A schematic depiction of the experiment
and algorithmic architecture of the analysis is depicted in Fig.~\ref{fig:rydberg} 

\subsection{Quantum error correction}
Quantum error correction (QEC) promises to help protect quantum information from
errors due to decoherence and quantum noise in quantum computation~\cite{devitt2013}. 
QEC is an essential ingredient for the future of fault-tolerant quantum information processing.
Machine learning techniques can help develop fast and flexible decoding algorithms for a wide
variety of quantum error correcting codes~\cite{PhysRevLett.119.030501,chamberland2018,PhysRevX.8.031084,sweke2018,Andreasson2019quantumerror,PhysRevLett.122.200501,PoulsenNautrup2019optimizingquantum,PhysRevResearch.1.033092}

Torlai and Melko~\cite{PhysRevLett.119.030501} first devised an algorithm for error 
correction in topological codes where the decoder is constructed from a RBM, as summarized
below. Ref.~\cite{PhysRevLett.119.030501} studies the two-dimensional toric code~\cite{kitaev2003a} and considers 
the simple phase-flip channel described by a Pauli operator where $\sigma_z$ is applied to each qubit with probability 
$p_{\text{err}}$. This operation produces error chains $\boldsymbol{e}$, whose boundary is called a
syndrome $\boldsymbol{S}(\boldsymbol{e})$ and can be accessed experimentally without destroying the 
quantum state. Error correction consists of applying a operator whose chain $\boldsymbol{r}$ generates 
the same syndrome. The recovery operation succeeds if the logical information in the code is 
unchanged by the operation. Datasets of error chains and their syndromes 
$\mathcal{D}=\{\boldsymbol{e},\boldsymbol{S} \}$ produced via numerical simulation. The dataset 
is used to train a model to approximate the underlying probability distribution $p_{\text{data}}(\boldsymbol{e},\boldsymbol{S})$ 
with a neural network. Once the model is trained, i.e., 
$p_{\text{model}}(\boldsymbol{e},\boldsymbol{S})\approx p_{\text{data}}(\boldsymbol{e},\boldsymbol{S})$,
the model takes an input syndrome $\boldsymbol{S}_0$ and samples the distribution 
$p_{\text{model}}(\boldsymbol{e}|\boldsymbol{S_0})$ until it generates an error  chain $\boldsymbol{e}_0$ 
compatible with $\boldsymbol{S_0}$. The resulting error chain $\boldsymbol{e}_0$ is selected for the 
recovery operation. Numerical results show that the neural decoder has a logical failure probability 
that is close to minimum weight perfect matching procedure~\cite{edmonds1965}. Building on these ideas, 
Ref.~\cite{chamberland2018} provides a summary of efforts on applying machine learning techniques to error 
correction, introduces several decoding algorithms based on deep neural decoders, and applies them to 
several fault-tolerant error correction protocols. Valenti and collaborators~\cite{PhysRevResearch.1.033092} 
introduce a neural-net-based approach to infer quantum Hamiltonians to aid the implementation 
of topological codes for quantum error correction. 

Reinforcement learning has also been applied to the  quantum error correction problem. 
Ref.~\cite{PhysRevX.8.031084} shows how an ``agent'' discovers quantum-error-correction strategies, 
protecting a collection of qubits against noise. This neural-network-based reinforcement 
learning approach constitutes a fully autonomous, human-guidance-free approach to the 
discovery of quantum error correction. Ref.~\cite{Andreasson2019quantumerror} implements 
a quantum error correction algorithm for bit-flip errors on the toric code using reinforcement learning and
finds that their algorithm is again close to the  minimum weight perfect matching algorithm for 
code distances up to $d=7$. Ref.~\cite{PoulsenNautrup2019optimizingquantum} presents a 
reinforcement learning framework for optimizing and adapting quantum error correction codes. 
The reinforcement learning algorithm learns to design good error correction codes that
make use of a small number of qubits. Ref.~\cite{sweke2018} introduces a reinforcement learning
framework for obtaining classes of decoding algorithms applicable to the fault-tolerant quantum
computation setting, which the authors exemplify utilizing deep Q-learning~\cite{mnih2015} 
to obtain surface code decoders for a variety of noise models.  

Finally, we mention that Ref.~\cite{PhysRevLett.122.200501} trained 
neural belief-propagation decoders for quantum low-density parity-check codes. The authors
report significant improvements for quantum low-density parity-check codes. Results
on the toric code, the quantum bicycle code, and the quantum hypergraph product code 
all show orders of magnitude of enhancement in decoding accuracy.  
 
In summary, the results in this subsection indicate that machine learning algorithms combined
with domain knowledge of quantum error correction represents a viable route for finding 
decoding schemes that perform on par with hand-made algorithms. These strategies open up the 
possibility to develop future machine learning decoders for more general error models and error 
correcting codes.  

\subsection{Quantum Control}

Quantum control,  the precise manipulation of physical systems whose behaviour is 
prescribed by the laws of quantum mechanics, has been a significant goal in quantum 
physics, chemistry, and engineering since the establishment of quantum mechanics~\cite{D'Alessandro:1251242}. 
Since optimal control theory lays the foundations of reinforcement learning algorithms 
to a large extent~\cite{10.5555/551283}, it is natural to expect that modern reinforcement learning technology 
can provide a platform for the accurate control of quantum systems. Ref.~\cite{PhysRevX.8.031086} 
implements a set of reinforcement learning algorithms and shows that their performance for the task
of finding short driving protocols from an initial quantum many-body state initial state 
to a target state is comparable to optimal control methods. The reinforcement learning methods developed in 
Ref.~\cite{PhysRevX.8.031086} use a single scalar reward, i.e., the fidelity of the 
state produced by the simulations of the physical with respect to the target state. In
a similar setting, Ref.~\cite{PhysRevLett.122.020601} provide convincing numerical evidence 
that such quantum control problems exhibit a universal spin-glass transition in the 
space of protocols as a function of the protocol duration. The authors suggest that the 
critical point exhibits a proliferation of protocols with nearly optimal fidelity, though 
the protocol with the truly optimal fidelity is exponentially hard to locate.   

In a similar vein, Ref.~\cite{niu2019} leverages the power of reinforcement learning to develop
a framework to optimize the speed and fidelity of quantum computation against leakage and 
stochastic control errors for a broad family of two-qubit unitary gates. The author's framework 
showcases an impressive improvement of two orders of magnitude in average-gate-error with respect 
to baseline stochastic gradient descent and up to a one-order-of-magnitude improvement 
in gate time from optimal gate synthesis counterparts. Similarly, Ref.~\cite{an2019} constructs 
a deep Q-learning framework to find the optimal time dependence of controllable parameters 
to implement a local Hadamard gate and a two-qubit CNOT gate. Importantly, Ref.~\cite{zhang2019b} 
benchmarks reinforcement learning for quantum control against traditional control methods for 
the problem of preparing a desired quantum state. More specifically, the authors compare the 
efficacy of three reinforcement learning algorithms, namely, tabular Q-learning, deep Q-learning, 
and policy gradient, with two traditional control methods: stochastic gradient descent and Krotov algorithms.
The authors find that the deep Q-learning and policy gradient algorithms outperform others techniques 
when the problem is discretized, namely, when the controls of the problem are discrete. These comparisons
shed light onto the suitability of reinforcement learning for quantum control problems. Finally, Ref.~\cite{youssry2020} 
introduced a deep learning framework based on recurrent neural networks for mitigating noise, 
and characterizing and controlling the dynamics of an open quantum system based on measurements.

Ref.~\cite{schafer2020} offers a complementary approach to the problem of quantum controls through 
the use of differentiable programming. The authors in Ref.~\cite{schafer2020} make use differentiable 
programming to backpropagate through the dynamical equations of the system and find optimal protocols
for state preparation of a single qubit, a chain of qubits, and a quantum parametric oscillator successfully.

\subsection{Quantum circuits and gates}

Variational quantum algorithms such as the variational quantum eigensolver (VQE)~\cite{peruzzo2014} or
the quantum approximate optimization algorithm (QAOA)~\cite{farhi2014} aim to simulate low-energy
properties of quantum many-body systems or to find approximate solutions of combinatorial optimization 
problems. These families of algorithms represent one of the most promising avenues for
observing computational advantages through near-term quantum computers. These algorithms 
employ quantum states produced by low-depth quantum circuits endowed with parameters that are
used to variationally optimize a cost function in the form of an expectation value of an operator
over the produced quantum state. While promising, these algorithms still face some significant 
challenges, and  in this context, machine learning techniques have been applied to several 
variational quantum state preparation problems. Ref.~\cite{arrazola2019} applies automatic differentiation 
to a differentiable photonic quantum computer simulator~\cite{Killoran2019strawberryfields} to 
find circuits of photonic quantum computers that perform a desired transformation between input and output states. 
Whereas in the case of a single input state the method discovers circuits for preparing a desired quantum state,
for the case of several input and output states, the method obtains circuits that reproduce the action 
of a target unitary transformation. Specific examples include learning short-depth circuits to 
synthesize single photons, cubic phase gates~\cite{ghose2007}, random unitaries, 
Gottesman–Kitaev–Preskill states~\cite{PhysRevA.64.012310}, NOON states~\cite{PhysRevA.40.2417}, and 
other states and gates.  

Ref.~\cite{verdon2019} considers the challenge of finding good parameter
initialization heuristics that improve convergence to a local minima of the parameterized circuit. 
The authors consider a meta-learning~\cite{vanschoren2019} approach where a classical 
neural network assists the learning of the quantum circuit. The neural network rapidly finds 
a global approximate optimum of the parameters, which is used as an initialization point for 
other local search heuristics. This combination yields superior optima of the quantum circuit, which
accelerates the search by several orders of magnitude with respect to other commonly used search strategies. 
Yao, Bukov, and Lin tackle a similar problem and show that policy-gradient-based reinforcement learning 
algorithms are well suited for the optimization of variational parameters of QAOA in a noise-robust 
fashion~\cite{yao2020}. Their technique is expected to help mitigate the unknown sources of errors in modern 
quantum simulators. An analysis of the performance of the algorithms in the context of quantum state transfer
problems demonstrates that excellent performance beyond state-of-the-art optimization algorithms. 

All in all, the studies described in this section suggest that machine learning technology can be successfully 
repurposed for controlling brittle quantum devices and quantum computers and can help with the design and 
improvement of variational algorithms in the era of approximate, near-term quantum computing. 
 
\section{Quantum physics inspired machine learning}
The connection between machine learning and physics has a long history beyond the recent adoption of machine
learning as a tool to study physical systems.  A prominent example of this connection is Hopfield's neural 
network model of associative memory~\cite{hopfield1982}. Hopfield's model consists of a single layer 
containing one or more fully connected recurrent neurons. The Hopfield network, which is commonly used for 
auto-association and optimization tasks, sparked  research about the application of spin glass theory to 
the understanding of neural networks~\cite{RevModPhys.91.045002}. 

The relation between physics and machine learning has experienced a recent revival with quantum 
systems inspiring new breeds of classical and quantum machine learning methods~\cite{NIPS2016_6211,Biamonte2017,gonzalez2020}, though
early suggestions of quantum neural networks abound~\cite{kak1995,behrman1996,behrman1999,schuld2014}. 
Here, we focus on classical machine learning methods inspired by quantum physics. In particular, 
we consider tensor networks, which originated in condensed matter theory and have served as a theoretical
tool to simulate and understand the role of entanglement in many-body physics~\cite{orus2019}. 
Stoudenmire and Schwab used the matrix product state representation~\cite{10.5555/2011832.2011833,orus2019}, 
also known as the tensor train decomposition in machine learning~\cite{oseledets2011}, for supervised 
learning, specifically for multi-class classification~\cite{NIPS2016_6211}. The strategy uses is based on
non-linear kernel learning, where input vectors $\boldsymbol{x}$ are mapped into a higher dimensional 
space via a non-linear function $\phi(\boldsymbol{x})$. This is followed by  a   decision function 
$f(\boldsymbol{x})=W\phi(\boldsymbol{x})$ which classifies the vector $\boldsymbol{x}$. The vector $W$ is 
a high-dimensional weight vector, which is, in turn,  approximated and optimized using tensor network methods. 
More precisely, the authors consider expressing $W$ as a matrix product state. The map $\phi(\boldsymbol{x})$ 
consists of a tensor product of the same local feature map applied to every component $x_i$ of the vector $\boldsymbol{x}$, 
i.e., $\phi(\boldsymbol{x})^{s_1,s_2,\ldots,s_N}=\phi(x_1)^{s_1} \otimes \phi(x_2)^{s_2} \otimes \ldots \otimes \phi(x_N)^{s_N}$, 
where $N$ is the dimensionality of the input vector $\boldsymbol{x}$ and the value $d$ is known as the local dimension. 
Therefore, each $x_j$ is mapped to a $d$-dimensional vector, and the full map can
be understood as a vector living in an $d^N$-dimensional space. The training of the model is based on a sweeping algorithm that      
sequentially optimizes a few of the local tensors of the matrix product state while leaving the rest fixed, in steps
resembling the expectation-maximization algorithm~\cite{Bishop2006}. Number-state preserving tensor networks, including
matrix product states, tree tensor networks~\cite{liu2019a}, as well as the multi-scale entanglement 
renormalization ansatz, have also been used for classification successfully~\cite{evenbly2019}.

Tensor networks have been successfully applied to several learning tasks including dimensionality reduction~\cite{cichocki2017},
unsupervised learning and generative modelling using matrix product states~\cite{PhysRevX.8.031012,stokes2019,bradley2019}, 
representation learning with multi-scale tensor networks~\cite{stoudenmire2018}, sequence-to-sequence learning
using matrix product operators~\cite{PhysRevE.98.042114}, language modelling~\cite{pestun2017,miller2020}, 
Bayesian inference~\cite{ran2020}. Ref.~\cite{NIPS2019_8429} provides a rigorous analysis of the expressive 
power of various tensor-network architectures for probabilistic modelling, including non-negative matrix product states
and Born machines~\cite{PhysRevA.98.062324}. Ref.~\cite{uranga2019} introduced a quantum inspired generative model suitable 
for raw audio signals. The architecture, which is equivalent to a continuous matrix product state~\cite{PhysRevLett.104.190405}, 
is built from a stochastic Schr{\"o}dinger equation describing the dynamics of a continuous time measurement applied on a quantum system.   
By construction, the model is autoregressive, which enables exact sampling of the distribution represented by the model. All these
developments have motivated the development of  a tensor-network library for physics and machine learning~\cite{roberts2019}.

Similarly, quantum circuits and quantum algorithms offer powerful approach toward the implementation of machine learning 
algorithms with the potential to exhibit quantum speedups~\cite{Biamonte2017,benedetti2019}. Here we focus on parameterized 
quantum circuits applied to data driven tasks, specifically supervised learning and generative modelling. Quantum parameterized
models have received widespread attention~\cite{benedetti2019} and have already been demonstrated experimentally in several experimental 
platforms\cite{havlicek2019b,otterbach2017,zhu2019,hu2019}. Their success has motivated the development of high level software 
tailored to the training of parameterized quantum circuits~\cite{bergholm2020,broughton2020}. Parameterized quantum circuits
are typically trained using a variational approach where a cost function is minimized with respect to the parameters in the circuit
so that  the converged model solves a computational problem. These algorithms encompass a mixture of quantum 
and classical computation where the quantum processor produces data to used to evaluate the cost function and its gradients, 
followed by a classical step where the parameters in the circuit are updated. A generic approach to supervised learning based
on quantum circuits, including classification and regression,  was  introduced in Ref.~\cite{PhysRevA.98.032309}. An important
ingredient clarified in Ref.~\cite{PhysRevA.98.032309}  was the estimation of gradients of cost functions over parameterized 
circuits for families of quantum gates based on exponential of Pauli strings in the context of supervised learning, which is now
used in several quantum machine learning proposals. Schuld and Killoran demonstrated that quantum computers can be used as a 
kernel method to efficiently perform computations in an intractably large Hilbert space and explored the theoretical foundations and the 
design of quantum machine learning algorithms based on such idea. Indeed, exploiting an exponentially large quantum state for 
machine learning has been demonstrated experimentally in Ref.~\cite{havlicek2019b}. Farhi and Neven~\cite{farhi2018} explore a 
parameterized quantum circuit that can represent labeled data, classical or quantum, and can be trained via supervised 
learning. The authors in Ref.~\cite{farhi2018} first consider the classification of classical data which consist of n-bit 
strings with binary labels and provide a practical example consisting of downsampled images of handwritten digits labeled 
as one of two distinct digits. In addition, the authors in Ref.~\cite{farhi2018} consider encoding the data as quantum 
superpositions of computational basis states corresponding to different label values and demonstrate numerically that 
training is possible. 

In the unsupervised learning setting, Verdon, Broughton and Biamonte~\cite{verdon2019a} introduced a low-depth variational 
quantum algorithm suitable for generative modelling tasks designed to sample low-energy distributions of Ising Hamiltonians.
An encompassing idea in unsupervised learning approaches is to think about the wavefunction produced with a quantum circuit 
as a generative model, an idea known as the quantum circuit Born machine~\cite{PhysRevA.98.062324}. Through numerical simulations, 
the authors in Ref.~\cite{PhysRevA.98.062324} explore an efficient gradient-based~\cite{PhysRevA.98.032309} learning algorithm for the 
Born machine by minimizing the kerneled maximum mean discrepancy loss. Their experiments demonstrate the importance 
of circuit depth and highlight the advantages of using gradient-based optimization. Ref.~\cite{benedetti2019a} proposes to 
use the performance on  generative modelling  tasks as a  benchmark  tool to characterize power of near-term quantum devices. 
The authors in Ref.~\cite{benedetti2019a} demonstrate their technique through numerical simulations and experiments on 
small ion trap quantum computers. Similarly, Ref.~\cite{zhu2019} experimentally demonstrates a quantum circuit training 
algorithm using bars-and-stripes dataset~\cite{10.5555/971143} and a quantum-classical hybrid machine based on trapped ions. 
Several extensions and training approaches for quantum generative modelling have been introduced, e.g., generative
adversarial quantum machines~\cite{PhysRevLett.121.040502,situ2020,PhysRevA.99.052306,romero2019},  where the key element is
to couple a quantum circuit generator and a classical or quantum discriminator which are trained 
simultaneously~\cite{goodfellow2014}. The learning is such that the generator tries to create statistics for data that 
mimics those of a dataset while the discriminator tries to discern true data from generated data. 

We conclude by briefly mentioning that quantum-inspired machine learning has drawn attention in areas beyond condensed matter
physics in areas such as high energy physics where there is an abundance of challenging big data problems~\cite{trenti2020}.

\section{Conclusions and outlook}
Modern machine learning techniques have started to spread through the landscape of quantum matter and strongly
correlated systems research. While cross-fertilization between physics and machine learning predates
the recent resurgence of applications to physical systems, the body of recent work reviewed here showcases the 
opportunities that machine learning techniques, ideas, and research culture 
can spark in the field of quantum many-body physics. A plausible goal for the near term is 
the development of models that combine quantum many-body physics with machine learning that deliver the necessary 
accuracy required for the prediction of novel phenomena at the speed of modern machine learning. Applied to
the different research areas discussed in this review, these developments may have the potential to help us find good 
approximations to open problems in quantum many-body systems including frustrated magnetism and fermionic matter 
in physical systems described by the Hubbard and the $t-J$ models~\cite{baeriswyl1995}. They may also significantly 
ameliorate the critical slowing down problem in classical and quantum Monte Carlo simulations~\cite{Newman1999}. Furthermore, 
these ideas may enable the simulation out-of-equilibrium dynamics of many-body systems beyond what is currently 
possible, and may ultimately help us delineate the boundary between quantum systems that can be simulated 
classically and those which would ultimately require quantum computing and quantum simulation strategies.

It is also natural to anticipate further contributions from physical sciences back to machine learning, as 
evidenced by the growing number of research studies related to the development and characterization of 
new physics-inspired machine learning models, their expressive power, and their training strategies. This 
include classical methods such as tensor networks but also quantum machine learning algorithms such as the 
quantum Boltzmann machine~\cite{PhysRevX.8.021050}, quantum Helmholtz machine~\cite{benedetti2018}, and the 
Born machine~\cite{PhysRevA.98.062324}. Likewise, tools from statistical mechanics have brought new conceptual 
advances to the field of deep learning where questions about expressiveness of deep neural networks, their 
information propagation capabilities, their generalization properties have been 
studied~\cite{bahri2020,RevModPhys.91.045002,NIPS2019_8921}. Tangentially, work a the intersection 
between physics and machine learning may aid interpretability more broadly since this has been 
a central topic the physics context~\cite{carrasquillaMachineLearningPhases2017,PhysRevE.98.022138,PhysRevLett.118.216401,zhang2019c,wetzel2020,PhysRevD.101.094507,dawid2020}.

Now is a privileged time for strongly correlated quantum systems research due to the enormous 
opportunities arising from artificial intelligence and quantum computing, two of today's most promising 
computing paradigms. Artificial intelligence is bustling with invigorating opportunities, research ideas, as well
as with research practices with great potential to impact computational and experimental physics research. 
We are only seeing the beginning of adoption of machine learning in the study of strongly correlated systems and quantum
matter. We anticipate that researchers in condensed matter, quantum information, atomic, molecular, and optical 
physics, and related areas of science will materialize this potential and produce exciting results 
through a sustained research effort at the intersection between machine learning and the broad area of quantum physics.

\section*{Acknowledgements}
I acknowledge Roger Melko, Giuseppe Carleo, Giacomo Torlai, Mohamed Hibat-Allah, Estelle Inack, Roeland Wiersema,
 Di Luo, Ehsan Khatami, Peter Broecker, Simon Trebst, Martin Ganahl, Isaac Tamblyn, Lauren Hayward, Peter Wittek, 
Ganapathy Baskaran, Bohdan Kulchytskyy, Francesco Ferrari, Federico Becca, Leandro Aolita, Felipe P\'erez, Maksims 
Volkovs, Ashley Milsted, Bryan Clark, Matt Beach, Kyle Cranmer, Evgeny Andriyash, Arash Vahdat, for collaborations 
and discussions related to machine learning and quantum physics.

\section*{Funding}
J.C. acknowledges support from Natural Sciences and Engineering Research Council of Canada (NSERC),  the Shared Hierarchical Academic Research Computing Network (SHARCNET), Compute Canada, Google Quantum Research Award, and the Canadian Institute for Advanced Research (CIFAR) AI chair program.

\section{References}


\end{document}